%% file: main.tex
\documentclass[sigconf]{acmart}

\usepackage{longtable}
\usepackage{tabularx}
\usepackage{enumitem}
\usepackage{csquotes}
\usepackage{todonotes}
\usepackage{array}
\usepackage{tabularray}
\usepackage{multirow}
\usepackage{geometry}
\newcolumntype{C}[1]{>{\centering\let\newline\\\arraybackslash\hspace{0pt}}m{#1}}
\newcolumntype{P}[1]{>{\centering\arraybackslash}p{#1}}

\graphicspath{{figures/}}

\usepackage{enumitem,varwidth, hyperref}
\usepackage{booktabs}
\usepackage{gensymb}

\usepackage{array}

\newenvironment{questionlist}
{\begin{enumerate}[label=\textbf{\arabic*.}, leftmargin=*, itemsep=0.2em, parsep=0.2em]}
  {\end{enumerate}}

\AtBeginDocument{%
  }

\copyrightyear{2025}
\acmYear{2025}
\setcopyright{cc}
\setcctype{by}
\acmConference[ASSETS '25]{The 27th International ACM SIGACCESS Conference on Computers and Accessibility}{October 26--29, 2025}{Denver, CO, USA}
\acmBooktitle{The 27th International ACM SIGACCESS Conference on Computers and Accessibility (ASSETS '25), October 26--29, 2025, Denver, CO, USA}
\acmDOI{10.1145/3663547.3746391}
\acmISBN{979-8-4007-0676-9/2025/10}

\begin{document}

\title{Characterizing Visual Intents for People with Low Vision through Eye Tracking}



\author{Ru Wang}
\email{ru.wang@wisc.edu}
\affiliation{%
  \institution{University of Wisconsin--Madison}
  \city{Madison}
  \state{WI}
  \country{USA}
}

\author{Ruijia Chen}
\email{ruijia.chen@wisc.edu}
\affiliation{%
  \institution{University of Wisconsin--Madison}
  \city{Madison}
  \state{WI}
  \country{USA}
}

\author{Anqiao Erica Cai}
\email{anqiaoc2@illinois.edu}
\affiliation{%
  \institution{University of Illinois Urbana-Champaign}
  \city{Champaign}
  \state{IL}
  \country{USA}
}

\author{Zhiyuan Li}
\email{zli2562@wisc.edu}
\affiliation{%
  \institution{University of Wisconsin--Madison}
  \city{Madison}
  \state{WI}
  \country{USA}
}

\author{Sanbrita Mondal}
\email{smondal4@wisc.edu}
\affiliation{%
  \institution{University of Wisconsin--Madison}
  \city{Madison}
  \state{WI}
  \country{USA}
}

\author{Yuhang Zhao}
\email{yuhang.zhao@cs.wisc.edu}
\affiliation{%
  \institution{University of Wisconsin--Madison}
  \city{Madison}
  \state{WI}
  \country{USA}
}


\renewcommand{\shortauthors}{Wang et al.}

\begin{abstract}

Accessing visual information is crucial yet challenging for people with low vision due to visual conditions like low visual acuity and limited visual fields. However, unlike blind people, low vision people have and prefer using their functional vision in daily tasks. Gaze patterns thus become an important indicator to uncover their visual challenges and intents, inspiring more adaptive visual support. We seek to deeply understand low vision users' gaze behaviors in different image-viewing tasks, characterizing typical visual intents and the unique gaze patterns exhibited by people with different low vision conditions.
We conducted a retrospective think-aloud study using eye tracking with 20 low vision participants and 20 sighted controls. Participants completed various image-viewing tasks and watched the playback of their gaze trajectories to reflect on their visual experiences. Based on the study, we derived a visual intent taxonomy with five visual intents characterized by participants' gaze behaviors. We demonstrated the difference between low vision and sighted participants' gaze behaviors and how visual ability affected low vision participants' gaze patterns across visual intents. Our findings underscore the importance of combining visual ability information, visual context, and eye tracking data in visual intent recognition, setting up a foundation for intent-aware assistive technologies for low vision people. 


\end{abstract}



\begin{teaserfigure}
  \includegraphics[width=\textwidth]{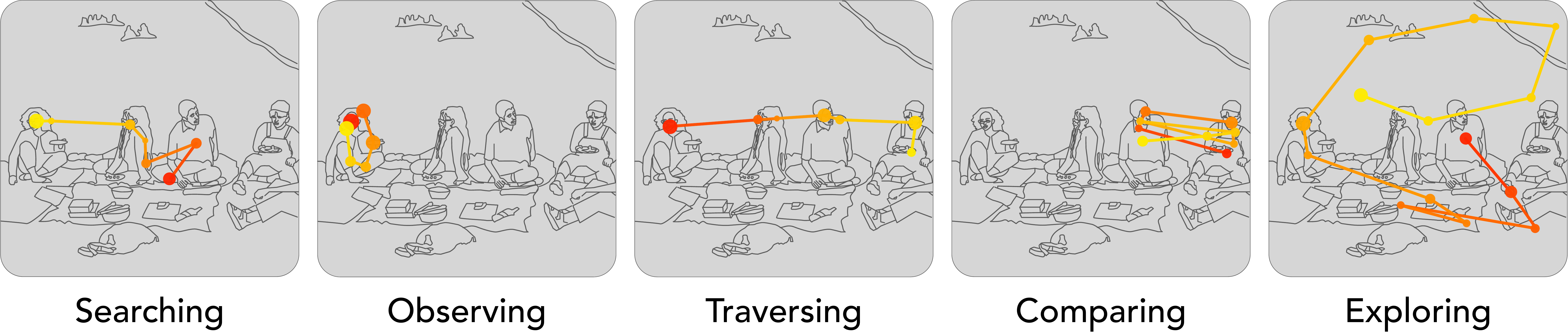}
  \caption{Overview of five visual intents---\textit{Searching}, \textit{Observing}, \textit{Traversing}, \textit{Comparing}, and \textit{Exploring}---identified through retrospective think-aloud study with both low vision and sighted participants using eye tracking. Each panel shows an illustration of the same example image, overlaid with a representative gaze trajectory for one visual intent. Gaze trajectories are color-coded to show progression (from red to yellow), and circles represent fixations, with size indicating fixation duration.}
  \Description{
   This figure shows a five-panel visual illustration showcasing five types of visual intent during image-viewing: Searching, Observing, Traversing, Comparing, and Exploring. Each panel contains an an illustration of the same example image (four people having a picnic on the foreground of the image) with mock gaze trajectories overlaid as colored lines and dots representing saccades and fixations. The "Searching" panel shows sparse gaze scanpaths moving towards the person on the left. "Observing" shows fixations clustered on the person on the left. "Traversing" displays scanpaths across multiple people one by one from left to right. "Comparing" includes back-and-forth gaze movements between the two people on the right. "Exploring" shows broader, scattered gaze paths covering the entire image. The gaze paths are illustrative and not drawn from real data.
  }
  \label{fig:teaser}
\end{teaserfigure}


\maketitle
\input{sections/1-introduction}

\input{sections/2-related_work} 
\input{sections/3-method}
\input{sections/4-results}
\input{sections/5-Discussion}

\begin{acks}
This research was supported in part by the National Science Foundation under Grant No. IIS-2147044 and the McPherson Eye Research Institute Grant Accelerator Program at the University of Wisconsin--Madison. We thank all our participants, as this work would not have been possible without their time and willingness to share their lived experiences.
\end{acks}

\bibliographystyle{ACM-Reference-Format}
\bibliography{sections/references}

\appendix
\renewcommand\thefigure{\thesection.\arabic{figure}}    
\setcounter{figure}{0}  
\renewcommand\thetable{\thesection.\arabic{table}}    
\setcounter{table}{0}  

\input{sections/appendix}

\end{document}

%% file: sections/1-introduction.tex
\section{Introduction}
Low vision is a visual impairment that cannot be fully corrected by eye glasses, contact lenses or other standard treatment \cite{nei}, affecting more than 2.2 billion people worldwide \cite{who_overview}. People with low vision experience diverse visual conditions such as central vision loss, peripheral vision loss, and blurry vision \cite{nei}, which significantly impact essential daily activities
\cite{szpiro2016people, ramulu2013difficulty, li2021non, wang2023practices, szpiro2016finding}. 

To tackle these challenges, vision enhancement technologies have been developed, such as optical and digital magnifiers that enlarge content details \cite{thomas2015assistive,hallett2018screen, maczoom}, edge enhancements to increase contrast \cite{holton2014review}, and image remapping tool to alleviate central vision loss \cite{deemer2018low}. While these tools can compensate certain visual difficulties, they provide limited support for essential visual tasks (e.g., reading) \cite{verghese2021eye, hallett2015reading, szpiro2016finding} due to their inability to adapt to user's visual context and goals \cite{szpiro2016finding, wang2024gazeprompt}. For example, when reading with a magnifier, a user's functional field of view is reduced as the entire reading material being magnified, making it difficult for low vision users to track reading positions \cite{verghese2021eye, hallett2015reading}, especially when switching lines \cite{wang2023understanding}. 
While researchers have designed task-specific low vision aids by considering users' surrounding environments \cite{zhao2016cuesee, ruan2024multi, zhao2019designing, fox2023using}, very few vision enhancement technologies have focused on detecting and supporting \textit{user intents}---the immediate goals behind visual behaviors that indicate users' dynamic needs in different tasks. To our knowledge, only one research by Wang et al. \cite{wang2024gazeprompt} incorporated user intent into low vision aids design.
However, this work only focused on reading tasks with relatively simple linear content structure and designed enhancements for only two predefined intents in reading context using rule-based algorithms.
To further push the boundary of assistive technology for low vision users, it is crucial to investigate low vision people's intents in more complex visual tasks (e.g., image-viewing), thus enabling more accurate and comprehensive intent recognition and inspiring intent-aware visual aids that provide more targeted and timely support.

Since eye gaze behaviors encode visual attention and mental states \cite{huang2015using}, eye tracking has been used broadly to understand sighted people's visual intents---distinct gaze patterns that reflect in-the-moment, meta-level objectives \cite{hu2021ehtask, david2021towards, malpica2023task, wang2024tasks, bednarik2012you, kiefer2013using, hild2018predicting}. 
However, prior work has typically leveraged predefined visual intents tied to specific tasks and contexts to train AI models for visual intent recognition. To date, no research has systematically identified and characterized a comprehensive set of visual intents across different visual tasks, which is essential for guiding the design of intent-aware technologies. 
Moreover, since low vision people's gaze behaviors can be significantly affected by their visual conditions \cite{wang2023understanding, heo2024reading, laude2018eye}, further investigation is needed to understand how visual intents are characterized by not only gaze behaviors but also visual abilities. 

Our research seeks to understand low vision users' gaze behaviors in image-viewing tasks, comprehensively identifying and characterizing their visual intents through \textit{a bottom-up, user-centered approach}. To achieve this goal, we conducted an eye-tracking-based retrospective think-aloud study \cite{cho2019eye, guan2006validity} with 20 low vision participants and 20 sighted controls, where they completed a series of image-viewing tasks and reflected on their gaze behaviors and challenges while watching the playback of their gaze trajectories. The image-viewing tasks were designed to elicit as diverse gaze behaviors as possible from participants to approximate their gaze behaviors in everyday life. With this study, we seek to derive a comprehensive taxonomy of visual intents for static content viewing, and investigate (1) how people's gaze behaviors vary across different visual intents, 
(2) how low vision people's gaze behaviors differ from those of sighted people under the same visual intents, and (3) how different visual abilities---specifically visual acuity and peripheral vision---affect low vision people's gaze behaviors under these visual intents.

Using both quantitative and qualitative methods, we 
identified and characterized five common visual intents shared by low vision and sighted participants---\textit{searching}, \textit{observing}, \textit{traversing}, \textit{comparing}, and \textit{exploring}. 
Our research showed that participants had different gaze behaviors under different visual intents. For example,  participants looked more at the background during searching and exploring than during other intents. We also discovered different gaze patterns between sighted and low vision participants under different visual intents, for example, low vision participants scanned more broadly than sighted participants during traversing and exploring.
Furthermore, we found both visual acuity and peripheral vision significantly shaped gaze behaviors across visual intents.
Based on our findings, we discuss suitable gaze data collection method for reliable visual intent inference, the generalizability of our visual intent taxonomy, more accessible visual intent recognition models that integrates visual ability and visual context, and design implications for intent-aware low vision assistive technologies.




%% file: sections/2-related_work.tex
\section{Related Work}

\subsection{Low Vision Experience \& Low Vision Aids}
Low vision is a visual impairment caused by different vision conditions such as cataract, macular degeneration, glaucoma, and many more eye diseases. These conditions can lead to low visual acuity, central vision loss and peripheral vision loss \cite{leat1999low} that cannot be fully corrected by eyeglasses, contact lenses or other standard treatments \cite{nei}. 
As such, low vision people experience various visual difficulties in different daily activities, such as reading \cite{legge1985psychophysics, szpiro2016people, tang2023screen, wang2023understanding},  navigating \cite{szpiro2016finding, zhao2018looks, tsuji12005landmarks}, cooking \cite{li2021non, wang2023characterizing, li2024contextual}, and socializing \cite{naraine2011social, rees2007self}.
Despite these challenges, people with low vision prefer completing daily tasks visually using their residual functional vision \cite{szpiro2016people}. 

To alleviate the impact of low vision conditions on daily tasks, various assistive tools and technologies have been developed, ranging from low tech tools (e.g., optical magnifiers\cite{virgili2018reading}) to high tech electronic devices (e.g., smartphone \cite{ioszoom, winmag}, head mounted display (HMD) \cite{zhao2015foresee, stearns2018design, zhao2016cuesee}). Magnification has been the most widely used visual support among people with low visual acuity and has been made available in mainstream personal devices as an accessibility feature, such as screen magnifiers on smartphones and laptops \cite{ioszoom, winmag}. Despite its broad use, magnification does not address visual field loss, for example, identifying letters can still be difficult for people with central vision loss even magnified \cite{wang2023understanding}. Furthermore, magnification can reduce users' usable field of view, leading to more visual challenges \cite{cheong2007relationship, hallett2015reading, ahn1995psychophysics, wang2023understanding, lee2007low}, such as tracking reading positions \cite{wang2023understanding}. Besides magnification, low vision people have also used other types of visual augmentations including contrast enhancement tools that use color filters to improve visibility of low contrast content \cite{zhao2015foresee, choudhury2010color, stearns2018design, ReBokeh}, contour enhancement technologies to enable clear separation of objects \cite{huang2019augmented, zhao2019seeingvr}, and color enhancement features that change certain colors to support people who are color blind \cite{tanuwidjaja2014chroma, langlotz2018chromaglasses}. In addition to visual clarity, researchers have developed technologies to support people with reduced peripheral vision by rendering a minified contour of a broader scene in the user's central vision on an HMD \cite{peli2006tunnelvision, zhao2019seeingvr}. Despite the effectiveness of these technologies, they are designed to alter a user's full vision uniformly without considering users' needs and context, which can lead to unnecessary visual distortion and distraction \cite{zhao2015foresee, kinateder2018using}.

Prior work has also investigated the design of low vision aids based on users' specific context and visual tasks, such as navigation \cite{huang2019augmented, zhao2020effectiveness, fox2023using, fox2023using, chen2025visimark}, visual search \cite{zhao2016cuesee, lang2021pressing}, cooking \cite{lee2024cookar}, and visual scanning \cite{jo2024watchcap}. For example, Fox et al. \cite{fox2023using} has explored the possibility of using different visual cues on a HMD to highlight obstacles for low vision users in indoor navigation. 
Lee et al. \cite{lee2024cookar} developed a head-mounted augmented reality (AR) system that recognized the affordances of kitchen tools---component parts that afford
interactions (e.g., kettle handles) and highlighted ``grabbable'' area of those tools to support safe and effective meal preparation. 
While prior research explored opportunity of providing targeted support for low vision users, they usually pre-specified a visual task and did not consider how to adapt to users' visual needs in-the-moment.

To our knowledge, GazePrompt \cite{wang2024gazeprompt} is the only work that incorporates gaze-based visual intents into low vision assistive technology---detecting line switching behaviors and word recognition difficulties and rendering visual and audio augmentations to improve reading experience. However, this work focused on relatively simple reading tasks and used rule-based methods to detect specific intents. No research has comprehensively investigated what visual intents low vision people exhibit in more complex scenarios. We seek to fill this gap by prompting and understanding low vision people's diverse visual intents in various image-viewing tasks.

\subsection{Recognizing Visual Intents}
As an important indicator of attention \cite{huang2015using} and information-gathering process \cite{lethaus2013comparison}, gaze behavior is key to uncovering low vision users' visual needs and challenges. Our goal is thus to understand these behaviors and characterize typical visual intents that can be recognized through gaze behavior. We define \textit{visual intents} as distinct gaze patterns that reflect in-the-moment, meta-level objectives. These visual intents are agnostic to specific \textit{visual tasks}, which refer to users' high-level goals. A user may employ multiple visual intents to accomplish a single visual task. For example, identifying people's activity in an image (visual task) could involve initially \textit{searching} for the people (visual intent) and subsequently \textit{comparing} their body poses to understand their interactions (another visual intent). In relation to other cognitive constructs, our definition of visual intent reflect overt visual attention \cite{geisler2011models} grounded in observable gaze behaviors. In contrast to intentionality, which refers to one's mental commitment to perform an action \cite{malle1997folk}, our visual intent specifically emphasizes the execution of that commitment via gaze.


While prior studies have explored visual intent recognition using eye tracking in various scenarios \cite{lethaus2013comparison, sattar2020deep, wang2024tasks, hu2021ehtask, yarbus2013eye}, there is no standard taxonomy of visual intents across visual tasks.
Instead, researchers have investigated how to recognize specific visual intents that are associated with certain visual tasks. 
For example, Bekta{\c{s}} et al.\cite{bektacs2023gear} defined \textit{reading}, \textit{inspecting}, and \textit{searching} in an AR-based robot repair task and evaluated the performance of different machine learning models on intent recognition (e.g., SVM). Hild et al. \cite{hild2018predicting} distinguished and recognized \textit{exploring}, \textit{observing}, \textit{searching}, and \textit{tracking} while participants watched street walking videos and achieved over 80\% accuracy using Random Forest classifiers on gaze data.
Lan et al. \cite{lan2022eyesyn} focused on museum-related visual intents like \textit{reading}, \textit{conversation}, \textit{browsing}, and \textit{watching}, and proposed a deep learning model that achieved up to 90\% accuracy through few-shot activity recognition using gaze data.

These studies demonstrated promising results in task-specific visual intent recognition, but their applicability remains limited to predefined tasks and constrained scenarios. There have been limited efforts on visual intent recognition in more generalizable scenarios, such as 360$\degree$ video watching in virtual reality (VR) \cite{hu2021ehtask, wang2024tasks}, which can potentially be applied to different real world scenarios. For example, Hu et al. \cite{hu2021ehtask} defined \textit{free viewing}, \textit{searching}, \textit{saliency estimation}, and \textit{tracking} visual intents in 360$\degree$ VR video viewing tasks and proposed a deep learning model, achieving high accuracy in both their dataset and a real world activity dataset. Despite the promising results, the visual intents were arbitrarily predefined without following any standard intent taxonomy. 

Besides the lack of comprehensive visual intent categorization, prior work all focused on sighted people who have different visual abilities, and thus gaze behaviors, from low vision people. As a result, critical research questions arise in intent-aware low vision technology: what are the visual intents exhibited by low vision users in visual tasks? What are the characteristics of their gaze behaviors under different visual intents to inspire accurate intent recognition?   
To address these questions, we seek to understand low vision people's visual intents by using an eye-tracking-based bottom-up approach that incorporates objective gaze data with users' subjective visual experience, thus deriving a comprehensive visual intent taxonomy that inspires timely and targeted intent-aware support for low vision people in various daily activities.

\subsection{Eye Tracking for People with Low Vision}
Although eye-tracking technology holds great potential for understanding human behaviors and intents, research on the eye movements of people with low vision remains limited. Researchers in optometry and vision science have investigated and compared the gaze behaviors of low vision and sighted people using eye tracking technology. Most research focused on examining detailed oculomotor behaviors, fixation stability, and binocular fixation in conditions including glaucoma \cite{lee2017effect, lee2019eye, senger2020saccadic, asfaw2018does,ganeshrao2021comparison}, Stardgart's disease \cite{giacomelli2021saccadic}, and age-related macular degeneration \cite{laude2018eye}. Some studies also investigated whether adjusting the location of the scotoma improves binocular contrast sensitivity \cite{alberti2017oculomotor} and whether enhancements in contrast and color improve visual comfort \cite{garric2021glaucoma} based on users' eye movements.

In the field of Human-Computer Interaction (HCI), little work has investigated how to leverage eye tracking technology to assist low vision people in daily tasks \cite{tang2023screen, maus2020gaze, schwarz2020developing, masnadi2020vriassist}. Schwarz et al. \cite{schwarz2020developing} designed a system to magnify what users were looking at on a computer screen. They tested it with three participants, but only one of them completed the calibration process successfully. Similarly, Maus et al. \cite{maus2020gaze} designed a gaze-guided magnification with a commercial eye tracker and found that five out of seven participants experienced a data loss of more than 50\%. Masnadi et al. \cite{masnadi2020vriassist} presented VRiAssist, which generated visual correction based on low vision users' gaze positions
to help them interact with virtual reality environments, but did not report detailed gaze data collection process. Wang et al. \cite{wang2023understanding} improved gaze calibration and data collection methods, enabling high quality gaze data collection from low vision users. They analyzed and compared low vision and sighted participants' gaze-level visual challenges in reading tasks, which led to the development of GazePrompt \cite{wang2024gazeprompt}, demonstrating the feasibility of gaze-based assistive technology for low vision users.

Though prior work has demonstrated early success in collecting high quality gaze data from low vision users and developing gaze-based low vision assistive technology, no prior work so far has thoroughly investigated gaze behavior patterns for people with low vision, which is a crucial step for understanding their visual intents and providing natural, targeted assistance. Contextualized in image-viewing scenarios, our research fills this gap by developing a taxonomy of visual intents and characterizing them using low vision people's gaze behaviors in various image-viewing tasks.

%% file: sections/3-method.tex
\section{Method}
Our goal is to comprehensively understand low vision people's visual intents and characterize them with gaze patterns and visual abilities. Unlike prior studies that predefined a set of visual intents based solely on eye movement measures (e.g., distance between fixations) \cite{wang2024tasks} or that catered specific scenarios such as map viewing \cite{kiefer2013using},
we use a user-centered bottom-up approach to comprehensively distill a visual intent taxonomy across various image-viewing tasks.

To achieve this goal, we collected low vision and sighted participants' gaze data during image-viewing tasks designed to stimulate diverse gaze behaviors and adopted a retrospective think-aloud study design \cite{cho2019eye, guan2006validity} to categorize their gaze behaviors into visual intents. Based on our visual intent taxonomy, we further conducted quantitative analysis to reveal the impact of visual abilities (sighted vs. low vision, low vs. high visual acuity, limited vs. intact peripheral vision) on participants' gaze behaviors. 



\subsection{Participants}
We recruited 20 low vision participants (P1-P20) and 20 sighted controls (C1-C20) for our study. Our low vision participants included 15 females, four males, and one non-binary, whose ages range from 21 to 82 ($Mean$ = 49.3, $SD$ = 20.3). Eight participants were legally blind. 
Participants had a wide range of visual conditions that are detailed in Table \ref{tab:lv_dem}. 
We recruited low vision participants from a local low vision clinic and via our university research email service. A participant was eligible if they were over 18 years old and had low vision. Participants were allowed to wear glasses unless they interfered with gaze calibration. Participants received compensations at \$20 per hour and were reimbursed for travel expenses. 

Our sighted participants included eight females and 12 males, with ages ranging from 19 to 71 ($Mean$ = 37.1, $SD$ = 12.3). They were recruited via our university research email service. All sighted participants' visual acuity in the better eye (corrected if with glasses) was no worse than 20/20 and had intact visual field. They were compensated \$10 per hour for participating in the study.

\input{sections/tables_figs/demo_table}

\subsection{Apparatus}

Our study was conducted in a well-lit lab. To streamline our study flow, we adopted a two screen setup, where participants were seated in front of a computer display (24-inch, 1920x1200 resolution) with a Tobii Pro Fusion (120Hz) eye tracker attached at the bottom (S1), and the researcher was seated before another screen with the same size and resolution (S2). The two screens were positioned so that participants could not see the researcher side (Fig \ref{fig:setup}a).

\subsubsection{Study Interfaces}
\label{interfaces}
We developed a gaze data collection and annotation system for our study. The system interfaces was developed with React \cite{react}. The gaze data were retrieved via the Tobii Pro SDK \cite{tobiiprosdk} in Python. We built a Flask-SocketIO \cite{flask-sio} server to process the gaze data and enable bi-directional communication between the server and the UI. We introduce the system interface below.

\begin{figure*}[h]
  \includegraphics[width=0.95\textwidth]{sections/tables_figs/gaze_ui.pdf}
  \caption{Our study set up and gaze data collection \& annotation system: (a) The positioning of  participant screen S1 and researcher screen S2. (b) An example of researcher interface on S2. (c) Control panel with segmentation and playback tools on S2. We show an example of gaze trajectory segment labeling. If a gaze trajectory segment (rectangular buttons below the slider) is selected, the corresponding gaze trajectory will be displayed. A gaze trajectory is represented by a sequence of circles with line segments in between them. Circles represent fixations, the longer the fixation duration, the larger the circle size.}
  \Description{This figure includes 3 subfigures---figure(a) on the top left corner of the parent figure, figure (b) on the top right, and figure (c) spanning the lower part of the parent figure. The figure overall shows our study set up and gaze playback \& annotation system. Figure (a) shows one screen S1 is positioned in front of the participant while another screen S2 is positioned in front of the researcher. The two screens are angled such that the participant cannot see S2. Figure (b) is an example of researcher interface on S2, where the researcher can see the participant's gaze trajectory in the middle, and control panel on the bottom of the screen. Figure (c) shows a closed up look of the control panel with segmentation and playback tools on S2, along with an example of gaze trajectory segment labeling. There are 4 gaze trajectory segments labeled Searching, Traversing, Exploring, and Comparing. Above each segment, there is an example showing the gaze trajectory segment overlay on a image. The four images are the same---a photo of a bus stop---but the gaze trajectory segments look different. A gaze trajectory (segment) is represented by a sequence of circles with line segments in between them. Circles represent fixations, the longer the fixation duration, the larger the circle size.}
  \label{fig:setup}
\end{figure*}

\begin{figure*}[h]
  \includegraphics[width=0.95\textwidth]{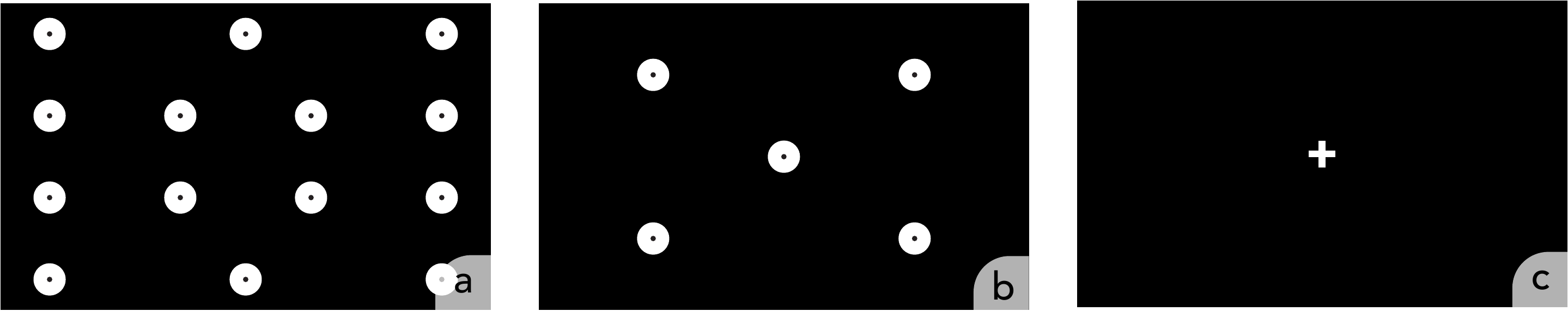}
  \caption{Our gaze calibration \& validation interface and visual field test interface: (a) 14-dot calibration. (b) 5-dot validation. (c) Visual field test with a fixation target at the center.}
  \Description{There are 3 figures showing the gaze calibration \& validation interface and visual field test interface. Figure (a) shows a 14-dot calibration interface in four rows. The first and fourth rows each contain three dots evenly spaced horizontally, while the second and third rows each contains 4 dots evenly spaced horizontally. Each dot is a white solid circle with a smaller black dot in the center.  Figure (b) shows a 5-dot validation interface, with one dot in the center and two dots aligned horizontally above the center dots and another two horizontally aligned dots below the center dot. Each dot is a white solid circle with a smaller black dot in the center. Figure (c) shows a visual field test interface with a fixation target---a plus sign---at the center.}
  \label{fig:cali}
    \vspace{-2ex}
\end{figure*}

\textit{\textbf{Gaze Calibration \& Validation.}} 
To collect high quality gaze data from low vision users, we adopted an accessible calibration process \cite{wang2023understanding}. We implemented a 14-dot gaze calibration (Fig \ref{fig:cali}a) and a 5-dot validation (Fig \ref{fig:cali}b) interface with resizable calibration targets following the configuration of prior work \cite{wang2024gazeprompt}. The default target size was 36px, and participants were allowed to adjust the target size to fit their visual ability.

\textit{\textbf{Visual Field Test.}} 
To estimate participants' on-screen visual field, we implemented a simplified visual field test with 44 dots surrounding the center of the screen based on guidance from prior work \cite{wang2023understanding} (Fig \ref{fig:cali}c). Participants were instructed to look straight ahead and press a key when a dot appeared in their peripheral visual field.

\textit{\textbf{Gaze Data Collection \& Annotation System}} 
On the researcher screen S2, the interface included a gaze display area---displaying the image selected during the study and participants' gaze trajectory overlay---and a control panel with segmentation and playback tools. 
The gaze trajectory was represented as a sequence of fixations (short pauses of gaze) generated from participants' raw gaze data using a dispersion-based real-time fixation detection algorithm \cite{kumar2008improving} to ensure fast processing speed during the study. The researcher could add and remove handles on the slider to adjust the length and number of gaze trajectory segments along the timeline and assign labels (i.e., visual intents) to them (Fig \ref{fig:setup}c). 
The researcher could play the entire gaze trajectory or a selected segment (Fig \ref{fig:setup}c). These features allowed the researcher to quickly understand participants' gaze trajectory and conveniently label visual intents for each segment. The color and size of the fixations in the gaze trajectory could be adjusted to cater each participant's preference and visual ability. The gaze playback interface on participant screen S1 only included a gaze display area which could be toggled on and off by the researcher. 

\subsubsection{Image Selection \& Question Design}
\label{img_questions} 
To ensure that our study covers a wide range of visual tasks and context to maximize the diversity of participants' gaze behaviors, we selected images aligned with participants' daily scenarios. Based on prior work on people's daily visual information needs \cite{kellar2006goal} and low vision people's visual challenges \cite{stangl2021going, gonzalez2024investigating}, we selected images from the following context: 1) \textit{News}, 2) \textit{E-commerce}, 3) \textit{Social Media}, 4) \textit{Travel}, and 5) \textit{Productivity}. 
Images were downloaded from context-dedicated websites (e.g., Amazon for e-commerce). To further diversify our image selection, within each context we included images with varying number of salient objects, varying number of object types (e.g., animal, plant), and images with and without a clear focal point. 

Furthermore, we designed questions tailored to each image to stimulate diverse gaze behaviors. Since directly optimizing for "diversity" is challenging, we adopted an alternative approach by designing questions that exhaust all possible levels of information in an image. Drawing insights from prior literature \cite{pandiani2023seeing, greisdorf2002modelling}, we categorized information in an image into the following three types (object can be a person): 1) \textit{Within-object information (level 1)}: including identification, details (e.g., color), and activity (e.g., body language) of an object ; 2) \textit{Cross-object information (level 2)}: including interaction or relationship between objects (e.g., the relationship between two people); 3) \textit{Overall interpretation (level 3)}: including the atmosphere of the image, the event the image describes. Our rationale was that by prompting participants to extract as many levels of information as possible, the diversity of their gaze behaviors could be maximized. Therefore, our questions were designed based on information levels. 

We collected 24 images for each context and removed politically sensitive ones, resulting in 117 images in total. All images were resized to be 1920x1200 and included three information levels, except 4 images with only two levels. For each image, we designed one question for each information level it had. During the study, each participant was presented six questions on six randomly selected images (excluding one tutorial trial), ensuring two questions per information level. The questions used in our study are listed in Appendix \ref{questionlist}.

\subsection{Procedure}
\label{procedure}
We conducted an eye-tracking-based retrospective think-aloud study. The study consisted of a single-session that lasted 1.5 to 2 hours. 

\textbf{\textit{Initial Interview \& Visual Ability Test.}}
We started with an interview covering participants' demographics after obtaining their consent. For low vision participants, we further asked about their visual conditions, daily visual difficulties, and experience with assistive technology. 
We then measured participants' visual acuity with wall-mounted letter-size ETDRS 1 and ETDRS 2 logMAR charts \cite{ferris1982new} for right eye and left eye, respectively. 
Participants were asked to sit at four feet from the eye chart, and for those who could not see the largest row, we tested their visual acuity at 2 feet. We further adjusted the distance for two participants who required more leg space. 
We tested their right eyes with chart 1 and left eyes with chart 2 (both with and without correction if they wore eyeglasses) by recording the smallest line they could recognize at least three out of five letters correctly without squinting. 
Our visual acuity test covered a range from 20/10 to 20/400 and we used self-reported visual acuity for those whose visual acuity was outside of the range of our test (Table \ref{tab:lv_dem}). 
We also measured participants' on-screen visual field using our simplified visual field test interface (Section\ref{interfaces}).

\textbf{\textit{Eye Tracking Calibration \& Validation.}}
We then conducted gaze calibration with participants with our customized calibration interface (Section \ref{interfaces}). Participants sat approximately 65 cm from a computer screen (S1) with an eye tracker. They were instructed to sit straight with their back touching the back of the chair to keep a stable and comfortable position throughout the study. We adjusted the screen height to align participants' eye level with the center of the screen. For low vision participants, we increased the calibration target size until they could locate the center of the target (the black dot) without squinting. They then completed 14-dot gaze calibration and 5-dot validation (Section \ref{interfaces}). We then collected participants' gaze data from the eye with better validation result in the following study phase if there was a significant difference in validation accuracy between the two eyes; otherwise, we used the average gaze position from both eyes.

\textbf{\textit{Retrospective Think-Aloud for Image-Viewing Tasks.}} 
After calibration, participants completed seven image-viewing trials in front of S1, with the first serving as a tutorial. The researcher sat in front of S2 (Fig \ref{fig:setup}a). This screen was used to monitor participants' gaze behavior and label visual intents on the fly. 

For each image-viewing trial, participants were first presented with a randomly selected question verbally (mapping to a certain viewing goal) before seeing the corresponding image. Only one question of that image was presented to each participant. 
When they were ready to proceed, the researcher would display the corresponding image and started collecting their gaze data. Participants were instructed to complete the visual task in their comfortable pace and press a key immediately to indicate the completion of the task. Afterwards, we asked participants to recall their gaze behaviors by describing where they looked at in the image in sequence, while the researcher played back the gaze trajectory only on S2 and preliminarily segmented their gaze trajectory based on their descriptions. Section \ref{interfaces} described the interface for gaze trajectory visualization and labeling. We chose this retrospective method to avoid any interruptions in participants' visual tasks that can cause gaze behavior changes. 

Next, to better categorize participants' gaze behaviors, we then showed the playback of the gaze trajectory to the participants on S1. The researcher adjusted the visualization of the gaze trajectory (e.g., color and size of the fixation circles) to ensure visibility to low vision participants. Participants were asked to further verify and explain their behaviors and intents (e.g., what they were doing) when viewing each labeled gaze trajectory segment.

Finally, the researcher refined the segmentation by merging segments with the same intent and splitting those with multiple intents according to participants' explanations. In total, our procedure made sure that each participant was presented six questions (order shuffled)---two level 1 questions, two level 2 questions, and two level 3 questions---from six randomly selected images from our image dataset (Section \ref{img_questions}). As a result, each participant contributed six segmented and labeled gaze recordings. 
Participants' responses were video recorded throughout the study for further analysis.

\subsection{Analysis}
We collected both qualitative and quantitative data. We describe our qualitative analysis first and then quantitative analysis.

\subsubsection{Generating Visual Intent Taxonomy}
\label{gen taxonomy}
We transcribed participants' video recordings of image-viewing tasks locally using Whisper model \cite{whisper} and manually corrected transcription errors.
We analyzed the transcript data using a standard qualitative analysis method \cite{saldana2021coding}. First, two researchers independently coded three sample transcripts from three participants using open coding, while watching the playback of their gaze trajectories to interpret and validate their responses. They developed an initial codebook, achieving substantial intercoder agreement (Cohen's Kappa = 0.73). They then discussed the codebook and resolved disagreements by jointly reviewing gaze replays to decide the best code for certain gaze behaviors, reaching full agreement on the codebook. The researchers then evenly split and coded the remaining transcripts, adding new codes to the codebook by mutual agreement. A third researcher oversaw the process to ensure a high-level agreement. Using affinity diagramming, the two researchers then collaboratively derived themes, including the visual intent taxonomy. 
Finally, they split all gaze segments and labeled each segment with the visual intent most closely matching the participant's reflection during the study. The researchers regularly checked each other's labeling and resolved disagreements. Unresolved gaze segments were discarded to ensure complete agreement on the final set of segments for subsequent analyses.

\subsubsection{Measuring Gaze Behavior}
\label{measures} 
We then characterized visual intents by investigating how gaze behaviors differed under different visual intents, between sighted and low vision people, and among different visual abilities (visual acuity and peripheral vision).
To further improve the accuracy of fixation detection from the raw gaze data, we used REMoDNaV \cite{dar2021remodnav}---an adaptive velocity-based eye movement event classification algorithm---to generate fixations for each gaze trajectory segment. We then manually corrected the fixation detection result by plotting the gaze coordinates (x and y) and velocity over time and checking for incorrect or missing fixation.
Our study resulted in 274 gaze trajectory segments for low vision participants and 236 gaze trajectory segments for sighted participants, with each segment assigned a visual intent category based on the process detailed in Section \ref{gen taxonomy}.
We specify the following measures for gaze behavior used in our analyses. 

(1) Low-level gaze measures: We analyzed the basic properties of gaze trajectories to characterize participants' gaze behavior at micro level: 
\begin{itemize}
    \item \textbf{Fixations:} Fixations are short pauses of gaze on an image to gather visual information \cite{rayner1998eye}. 
    For each participant and visual intent, we computed the \textbf{mean fixation duration} and \textbf{fixation rate} (number of fixations per second).
    \item \textbf{Saccades:} Saccades are the rapid eye movements connecting two consecutive fixations \cite{mahanama2022eye}. 
    We measured \textbf{mean saccade amplitude}, defined as the distance between two consecutive fixations, averaged for each visual intent for each participant.
\end{itemize}

(2) Spatial gaze measures: To characterize participants' gaze behavior globally, we sought to understand how participants shifted their visual attention among different areas of an image. To this end, we divided the image-viewing area into 8x5 grids, resulting in 40 area of interest (AOIs). We introduce the following measures:
\begin{itemize}
    \item \textbf{Stationary Entropy:} To investigate the spatial dispersion of fixations in a gaze trajectory, we calculated the stationary entropy for each gaze trajectory segment as $H_s = - \sum_{i}\pi_i \log_2 \pi_i$ \cite{mahanama2022eye} where $i$ indicates the index of AOI, and $\pi_i$ means the observed probability of fixation landing in the $i$th AOI. 
    A low $H_s$ indicates that visual attention is concentrated towards certain AOIs, whereas a high $H_s$ indicates more equally distributed visual attention across all AOIs. We computed this measure for each trajectory segment and average it across all instances of each visual intent that a participant performed.
\end{itemize}

(3) Context-aware gaze measures: Since gaze behaviors are tied to image content, we characterized participants' gaze behaviors based on their interaction with semantically meaningful areas. To support this analysis, two researchers annotated segmentation masks for \textit{salient objects} in each image using outputs from the Segment Anything Model 2 (SAM 2) \cite{ravi2024sam2} on Roboflow \cite{dwyer2024roboflow}, and reconciled discrepancies to finalize the annotations. Based on these annotated objects, we define the following measures below for this analysis:
\begin{itemize}
    \item \textbf{Number of Objects Visited:} To evaluate the breadth of a person's gaze behavior, we counted how many distinct objects a participant fixated during each gaze trajectory segment. A fixation was considered to land on an object if a circle centered at the fixation point---with a radius equal to the eye tracker's validation error for that participant---intersected the segmentation mask of the object. This was to compensate the eye tracking error to further improve the accuracy of our measure. We computed this measure for each visual intent a participant performed.

    \item \textbf{Attention Distribution over Objects:} To characterize how participants distributed their visual attention across salient objects in an image, we calculated the \textbf{object attention variability}---standard deviation of the time spent on each object within a given image---where higher values indicate more uneven attention allocation. Additionally, we computed the \textbf{foreground attention ratio}, defined as the proportion of time spent on all salient objects, to assess the overall focus on image foreground versus background. Together, these measures provided a more nuanced understanding of participants' gaze behavior at the object level. For both measures, we took the average across gaze trajectory segments within each visual intent for each participant.

\end{itemize}

\subsubsection{Comparing Gaze Behavior between Low Vision and Sighted People}
\label{ana: s vs lv}
To understand how low vision and sighted people's gaze behavior differ under different visual intents, we define a within-subject factor \textit{\textbf{VisualIntent}} and between-subject factor \textbf{\textit{Vision}}.

Applying the result of our visual intent taxonomy (detailed in Section \ref{taxonomy}), \textit{\textbf{VisualIntent}} included five levels---\textit{Searching}, \textit{Observing}, \textit{Traversing}, \textit{Comparing}, and \textit{Exploring}---shared by both low vision and sighted participants. Factor \textbf{\textit{Vision}} included two levels---\textit{Sighted} and \textit{LowVision}. All measures defined in Section \ref{measures} were involved in this analysis. We first checked the normality of each measure using Shapiro-Wilk test. If normally distributed, we fitted our data with Linear Mixed-Effects (LME) Model and calculated the ANOVA table for p-values for the fixed effects \cite{kuznetsova2017lmertest}; Tukey's HSD was then used for post-hoc comparison if significance was found. Otherwise, we used Aligned Rank Transform (ART) ANOVA and ART contrast test for post-hoc comparison \cite{elkin2021aligned}. We used partial eta squared ($\eta_p^2$) to indicate effect size, with 0.01, 0.06, 0.14 representing the thresholds of small, medium and large effects \cite{cohen2013statistical}. 

\subsubsection{Investigating the Effect of Visual Abilities on Low Vision People's Gaze Behavior}
For low vision participants specifically, we compared the gaze behavior under different visual intents for people with different visual abilities. 
Similarly, we had one within-subject factor \textit{\textbf{VisualIntent}} including five levels described in Section \ref{ana: s vs lv}. For visual abilities, we involved two between-subject factors: \textit{\textbf{PeripheralVision}} including two levels---\textit{Limited}, \textit{Intact}---based on participants' self-report and our simplified visual field test, and \textit{\textbf{VisualAcuity}} with two levels---\textit{Low}, \textit{High}---with 20/100 in the better eye as threshold \cite{wang2024gazeprompt}. We adopted the same statistical analysis detailed in Section \ref{ana: s vs lv}.

%% file: sections/tables_figs/demo_table.tex
\begin{table*}[h]
\scriptsize
\centering
\begin{tabular}
{C{0.4cm}C{0.5cm}C{3cm}C{0.7cm}C{1cm}C{2.2cm}C{3.4cm}C{3.4cm}}
\toprule
\textbf{ID}&\textbf{Age/{\newline}Gender}&\textbf{Diagnosed{\newline}Condition}&\textbf{Legally {\newline} Blind?} &\textbf{Visual Acuity}&\textbf{Visual Field}&\textbf{Other Visual {\newline} Difficulties}&\textbf{Accessibility {\newline} Tech Used}\\
\midrule
P1 & 72/M & Macular degeneration & N & L: 20/320{\newline} R: 20/400 & Central vision loss & N/A & Large font, invert color, screen magnifier \\
\hline
P2 & 62/F & Spinal meningitis & Y & L: 20/2200{\newline} R: 20/320 & Peripheral vision loss & N/A & Full-screen magnifier, Large font \\
\hline
P3 & 58/F & Retinitis pigmentosa & Y & L: 20/160{\newline} R: 20/120 & Peripheral vision loss & Color blind; light sensitive & Brighter and Bigger, large font, screen magnifier, invert color \\
\hline
P4 & 31/F & Retinitis pigmentosa & N & L: 20/25{\newline} R: 20/25 & Peripheral vision loss & N/A & Large and bold font, night-time mode, \\
\hline
P5 & 72/F & Macular degeneration & Y & L: \textless{}20/400{\newline} R: 20/100 & Central vision loss & Colors appear darker; light sensitive & Text-to-speech, screen magnifier \\
\hline
P6 & 72/F & Cone dystrophy & Y & L: 20/160{\newline} R: 20/160 & Central vision loss & Difficulty with black and navy blue, orange and pink & Screen magnifier, large font \\
\hline
P7 & 41/M & Retinitis pigmentosa & Y & L: 20/30{\newline} R: 20/30 & Peripheral vision loss & Light sensitive & Large font and pointer \\
\hline
P8 & 31/F & Congenital glaucoma & N & L: \textless{}20/400{\newline} R: 20/125 & Peripheral vision loss & Difficulty with darker shades; light sensitive & Large and bold font, invert color \\
\hline
P9 & 35/F & Retina detachment, puckered macula (left) & Y & L: 20/400{\newline} R: blind & Peripheral vision loss & Difficulty with purple and blue; light sensitive & Text-to-speech, dark mode, large font, screen magnifier \\
\hline
P10 & 64/F & Diabetic retinopathy, glaucoma, cataract & N & L: 20/84{\newline} R: 20/84 & Peripheral vision loss & Light sensitive & Screen magnifier \\
\hline
P11 & 77/F & Glaucoma, ICE syndrome (right) & N & L: 20/30{\newline} R: \textless 20/400 & Slight peripheral vision loss & Light sensitive & Zooming in \\
\hline
P12 & 57/F & Retinitis pigmentosa & Y & L: 20/100{\newline} R: 20/100 & Peripheral vision loss & Difficulty with dark green and dark blue, yellow and green; light sensitive & Zooming in, large font \\
\hline
P13 & 62/F & Scar tissue on retina & N & L: 20/160{\newline} R: 20/96 & Central vision loss (right), peripheral vision loss & Light sensitive & Invert color, screen magnifier \\
\hline
P14 & 29/M & Knobloch syndrome & Y & L: \textless{}20/400{\newline} R: 20/200 & Peripheral vision loss & N/A & Screen magnifier, large font, dark mode \\
\hline
P15 & 28/M & Stargardt's disease                                                            & N & L: 20/96 {\newline} R: 20/96 & Slight central vision loss      & Light sensitive                     & Screen magnifier                                                         \\
\hline
P16 & 29/N & Cone-rod dystrophy                                                             & N & L: 20/40 {\newline} R: 20/70   & Central vision loss            & Light sensitive; slight color blind & Color filter, invert color, screen magnifier, large font, text-to-speech \\
\hline
P17 & 21/F & Stargardt's disease                                                            & N & L: 20/60 {\newline} R: 20/125  & Central vision loss            & Light sensitive  & Large font, high brightness                                              \\
\hline
P18 & 37/F & Congenital posterior subcapsular polar cataract, posterior vitreous detachment & N & L: 20/50 {\newline} R: 20/50   & Intact                         & Light sensitive                     & Screen magnifier, invert color                                           \\
\hline
P19 & 25/F & Nystagmus                                                                      & N & L: 20/80 {\newline} R: 20/80   & Intact                         & Slight color blind                     & Screen magnifier                                                         \\
\hline
P20 & 82/F & Macular degeneration                                                           & N & L: 20/200 {\newline} R: 20/80  & Central vision loss (left) & Light sensitive       & Screen magnifier                       \\                                 

\bottomrule
\end{tabular}
\caption{Demographic information of 20 low vision participants. The visual acuity of P4, P7, P11, P12, P15 were post-correction since they wore glasses during the study. }
\label{tab:lv_dem}
  \vspace{-8ex}

\end{table*}

%% file: sections/4-results.tex
\section{Results}
We first report the quality of gaze data collected from low vision and sighted participants to establish the reliability of our subsequent analyses. We then introduce our visual intent taxonomy and highlight the unique gaze patterns exhibited by low vision participants. Next, we present quantitative comparisons of gaze behaviors across different visual intents and between low vision and sighted participants. Finally, we examine how visual abilities (visual acuity and peripheral vision) affect gaze behaviors among low vision participants under each visual intent.

\subsection{Gaze Data Quality}
The mean angular errors from the 5-dot validation was 0.99$\degree$ (SD = 0.62$\degree$), or approximately 45 pixels, for low vision participants, and 0.53$\degree$ (SD = 0.14$\degree$), about 24 pixels, for sighted participants. Low vision participants exhibited higher errors than sighted participants, consistent with prior work \cite{wang2023characterizing}. However, the errors were substantially smaller than the average height and width of salient objects in our image stimuli (over 400 pixels), indicating sufficient eye tracking accuracy for interpreting gaze behavior in relation to image content. Mean gaze data loss---the proportion of invalid (lost) gaze points to the total collected---was 7.6\% for low vision participants, and 2.3\% for sighted participants. Both values fall within the range of usable data loss rate in prior work \cite{cuve2022validation}. These results confirm that the gaze data quality was adequate for both participant groups, supporting the validity of our subsequent analysis on visual intents and gaze behaviors.

\subsection{Visual Intent Taxonomy}
\label{taxonomy}
By triangulating participants' gaze recordings with their subjective visual experiences, we found low vision and sighted participants shared five common visual intents---\textit{searching}, \textit{observing}, \textit{comparing}, \textit{traversing}, and \textit{exploring}. However, we noticed low vision participants demonstrated unique goals and gaze behaviors beyond those visual intents, rooted from their unique visual experience. We characterize them below (Fig \ref{fig:taxonomy}): 

\begin{figure*}[h]
  \includegraphics[width=\textwidth]{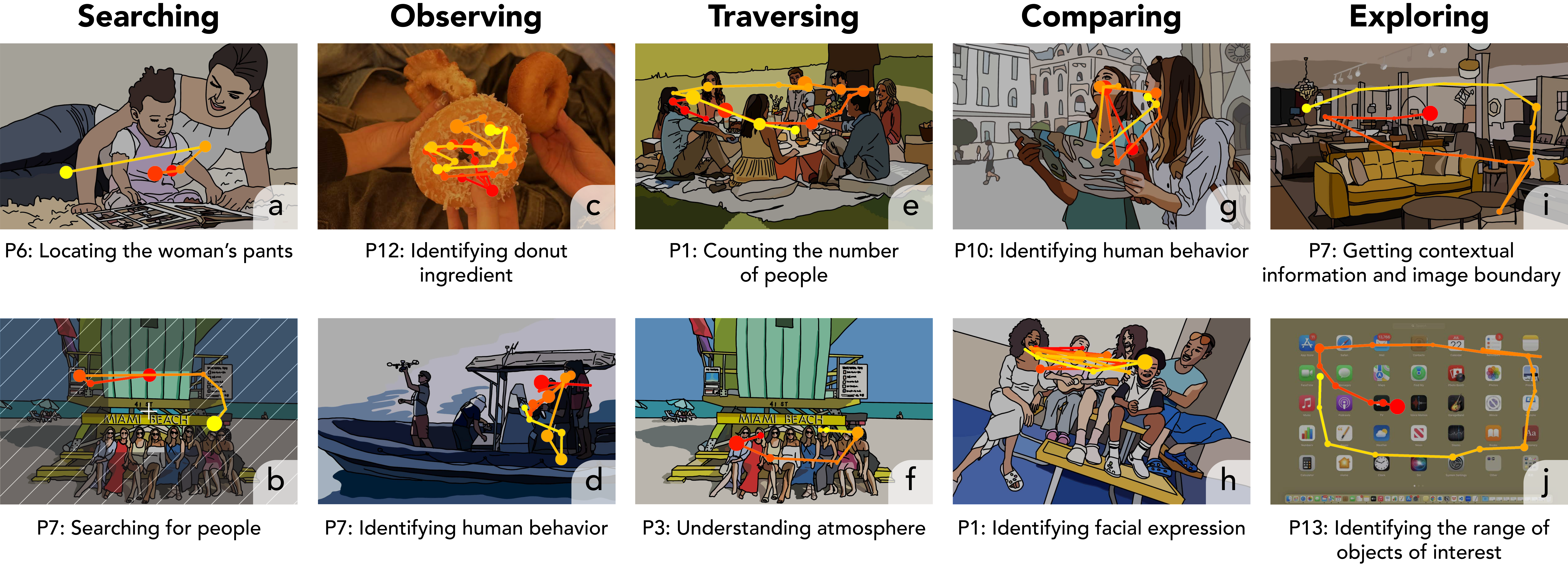}
  \caption{Examples of different visual intents exhibited by low vision participants. Gaze trajectory overlay on the images are presented as a sequence of fixations (circles) and saccades (line segments). The gaze trajectory is color-coded to show progression, transitioning gradually from red (starting point) to yellow (endpoint). The size of circle represents fixation duration. The examples in the same column share the same visual intent but for different tasks. Note: For image stimuli with copyright restrictions, we present author-created illustrations to represent the layout and key elements of those images.}
  \Description{This figure presents examples for different visual intents. Gaze trajectory overlay on the images are presented as a sequence of fixations (circles) and saccades (line segments). The gaze trajectory is color-coded to show progression, transitioning gradually from red (starting point) to yellow (endpoint). The size of circle represents fixation duration. The figure consists of two rows and five columns: each column represents a visual intent---searching, observing, traversing, comparing, and exploring (from left to right)---while examples in the same column share the same visual intent but for different tasks. Column 1: Searching. Picture a shows a mom and a child reading, with brief gaze trajectory overlay that represent searching. There are text below the image: ``P6: Locating the woman's pants''. Picture b depicts 8 people posing before a camera on the beach, with a gaze trajectory overlay and a visual field overlay, showing that the searching gaze trajectory was moving along the boundary of searching. There is a line of text below the image: ``P7: searching for people''. Column 2: Observing. Picture c shows 3 donuts, with gaze trajectory overlay that represent observing. below the picture, a line of text: ``P12: identifying donut ingredient''. Picture d shows a boat with 3 people on it, with gaze trajectory overlay that represent observing on two people on the right. below the picture, a line of text: ``P7: identifying human behavior.'' Column 3: Traversing. Picture e shows a picnic scene where people sit around a table, with gaze trajectory overlay that represent traversing through the people one by one. below the picture, a line of text: ``P1: counting the number of people''. Picture f is the same image as picture b, but with gaze trajectory overlay that represent traversing through the people one by one. below the picture, a line of text: `` P3: understanding atmosphere.'' Column 4: Comparing. Picture g shows 2 people laughing with a map on hand, with gaze trajectory overlay that represent comparing. below the picture, a line of text: ``P10: identifying human behavior.'' Picture h shows 5 people sitting on a bench smiling, with gaze trajectory overlay that represent comparing. below the picture, a line of text: ``P1: identifying facial expression.'' Column 5: Exploring. Picture i shows a furniture store, with gaze trajectory overlay that represent exploring. below the picture, a line of text: ``P7: getting contextual information and image boundary.'' Picture j shows a screenshot of macOS full of app icons, with gaze trajectory overlay that represent exploring. below the picture, a line of text: ``P13: identifying the range of objects of interest.''}
  \label{fig:taxonomy}
  \vspace{-2ex}
\end{figure*}

\textit{\textbf{Searching}} is characterized by a sequence of fixations directed towards a target object. During searching, participants actively assessed whether an object was relevant and decided whether to continue searching. Fig \ref{fig:taxonomy}a shows how P6 directed her gaze to the woman's face and then located her pants during a task of finding the woman's pants. Searching was also used to confirm the absence of objects, ensuring no further searching was needed (four low vision and nine sighted participants). We notice that participants with peripheral vision loss searched along the boundary of their visual field to identify objects they otherwise might miss in their peripheral vision. P7 explained that he actively scanned areas beyond his visual field to gather information, as illustrated in Fig \ref{fig:taxonomy}b, where he directed his gaze to the left to see objects originally outside of his visual field. 




\textit{\textbf{Observing}} is defined as a sequence of fixations primarily concentrated on a single object to identify its identity (e.g., type or name) (16 low vision and nine sighted) and details (e.g., color, texture) (12 low vision and seven sighted) or on a single person to identify their activity or facial expressions (eight low vision and 14 sighted). Fig \ref{fig:taxonomy}c shows an example of P12 observing a donut to identify the ingredients on top of it.
For participants with peripheral vision loss (P2, P7), observing looked a bit different than others---they tended to fixate near the outline of the object(s) to ensure they had acquired all information about this object. Fig \ref{fig:taxonomy}d shows an example where P7 observed the behavior of two people on the right side of the image while ensuring no contextual information was overlooked to compensate for his limited peripheral vision. Unlike other participants, instead of examining salient objects in an image, some low vision participants also observed hidden cues in some visual tasks. P2 observed the lower part of an image stimulus (curb and side walk) as the first step of determining the location the photo was taken. As she explained, \textit{``I was looking for distinctions like a sidewalk or sand on a beach... To look for a sidewalk or a road or you know cornfield or a beach or steps.''}

\textit{\textbf{Traversing}} is characterized by a sequence of fixations across adjacent objects or text, often performed for tasks such as counting (13 low vision and 18 sighted) and reading (two low vision and six sighted). Fig \ref{fig:taxonomy}e demonstrates how P1 counted the number of people in the image by fixating on each person in a counter-clockwise order.
In addition, traversing was used to understand the collective context of a scene, such as examining objects in the image one by one to infer group activities and object relationships (five low vision) or the overall atmosphere (P15). 
Fig \ref{fig:taxonomy}f illustrates how P3 interpreted the emotion the image conveyed by traversing over people's clothing and faces.

\textit{\textbf{Comparing}} is defined as a series of fixations shifting back and forth between two or more objects to identify relationships or dynamics between them (11 low vision, 14 sighted). For example, Fig \ref{fig:taxonomy}g shows how P10's fixations jumped between the map and two people's faces to connect them and infer their behavior.
Participants also compared size of objects (P4, P7, C5), decided whether multiple objects belong to the same category (five low vision and two sighted), and compared people's facial expression to infer emotion of the image (eight low vision and two sighted). Unlike sighted participants, we found some low vision participants (P1, P2, P11) compared nearby objects or people merely to gather additional visual context when direct observation of the target object was challenging. In Fig \ref{fig:taxonomy}h, P1 was unsure if the woman on the left was smiling due to his central vision loss. Rather than continuing to observe the target person, he examined nearby people's faces and compared their facial expressions to gain confidence in his judgment, as he commented, \textit{``What I see is closer here, these people (on the right of the woman) are smiling, so I'm guessing she's smiling.''} 


\textit{\textbf{Exploring}} is characterized by widely distributed fixations across the entire image to gather a variety of contextual information, such as the location (nine low vision and two sighted), salient objects (10 low vision, 11 sighted), and people's activities (three low vision, six sighted), anything relevant to the visual task they performed.
For participants with limited peripheral vision (P7, P8, P11), exploring was also used to roughly identify and constrain the range of the potential area of interest for subsequent visual intents. Fig \ref{fig:taxonomy}i illustrates an example of P7 carefully exploring what is in the image while identifying the boundary of the image, as he elucidated, \textit{``I'm like kind of center here and as I work my way around I wanted to kind of go down to see where the bottom of the screen was to make sure I wasn't missing anything below. And then I work my way up to where I can see the top of the screen.''} For the same reason, we found participants with peripheral vision loss tended to explore in a more systematic way---they often started from the center of the screen and spiraled outward to minimize the risk of missing information (P2, P3, P7, P12) (Fig \ref{fig:taxonomy}i, \ref{fig:taxonomy}j).
For participants with intact peripheral vision, they could rely on their peripheral vision to gather contextual information rather quickly without needing to fixate across the entire image.

\begin{figure*}[h]
  \includegraphics[width=\textwidth]{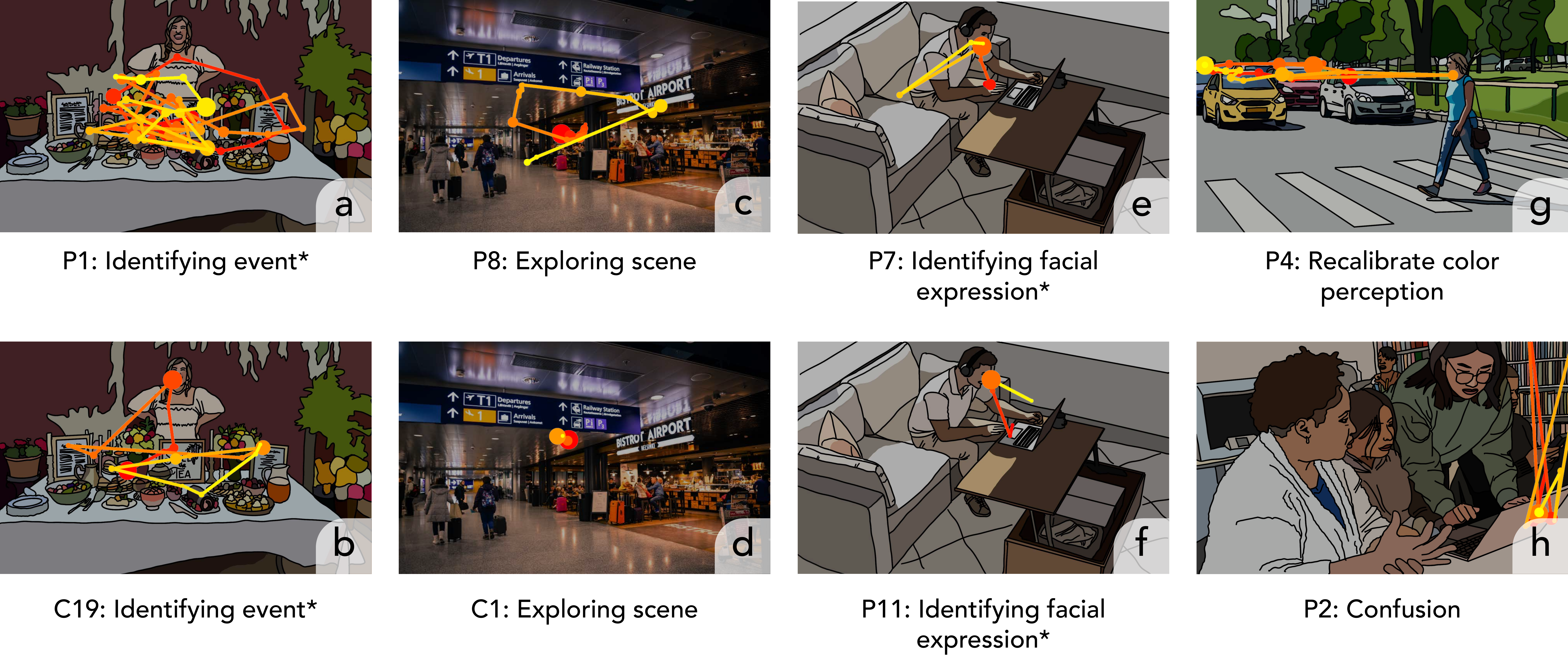}
  \caption{Examples of gaze behaviors demonstrated by our participants, influenced by visual intent, visual condition (low vision vs. sighted), and visual ability. From \textit{a} to \textit{f}, the images with gaze trajectory overlay on the same column represent the same visual task. * means the gaze trajectory shown on the image were the full gaze trajectory, not specific visual intent segments. Picture \textit{g} and \textit{h} represents two unique gaze behaviors demonstrated by low vision participants. Note: For image stimuli with copyright restrictions, we present author-created illustrations to represent the layout and key elements of those images.}
  \Description{This figure presents examples of gaze behaviors affected by visual intent, visual condition (low vision vs. sighted), and visual ability. The figure consists of two rows and four columns, and the images with gaze trajectory overlay on the same column (a and b, c and d, e and f, g and h) represent same task. Picture (a) shows a table full of food and a lady standing behind it, with gaze overlay scattering around the food on the table. there is a line of text below: ``P1 identifying event with a star (*) marker'', meaning the gaze trajectory shown on the image were the full gaze trajectory, not specific intent segments. Picture (b) shows the same photo as (a), but with gaze overlay sparsely around the food on the table. there is a line of text below: ``C19 identifying event with a star (*) marker''. Picture (c) is a photo taken inside an airport, with scattered fixations around the middle of the image representing exploring. there is a line of text below: ``P8 exploring the scene.'' Picture (d) shows a same image with (c), with few fixations concentrated in the center of the image. there is a line of text below: ``C1 exploring scene.'' Picture (e) shows man sitting on a sofa using a laptop, with gaze overlay crossing the man and briefly spilling over to the sofa. there is a line of text below: ``P7 identifying facial expression with a star (*) marker''. Picture (f) shows the same picture as (e) with gaze overlay crossing the man and the region just in front of the man. there is a line of text below: ``P11 identifying facial expression with a star (*) marker.'' Picture (g) shows a lady walking on a cross walk with cars stopping, with gaze overlay jumping between the cars and the lady. there is a line of text below: ``P4 recalibrating contrast perception.'' Picture (h) shows 3 people in a library with a laptop on the desk, with a gaze overlay on the right side of the image going beyond the image boundary. there is a line of text below: ``P2  confusion.''}
  \label{fig:stats}
  \vspace{-2ex}
\end{figure*}

\textit{\textbf{Unique Gaze Behavior from Low Vision Participants.}}
Furthermore, due to specific visual conditions, low vision participants demonstrated unique visual behaviors beyond the above five visual intents. When interacting with low contrast image content, some participants occasionally directed their gaze to areas with higher contrast or better visibility to `restore' their confidence in perceiving colors accurately (P4, P17).
In Fig \ref{fig:stats}g, while counting the number of cars with similar colors on the background, P4 briefly shifted her gaze to the woman in the foreground because it was more salient and had higher contrast. 
By doing this, she could recalibrate her color perception and feel more confident about her ability to correctly count the cars in the low contrast area. As she explained, \textit{``Sometimes... I focus on something that's easier for me to see and then go back to, you know, like trying to answer the question, cause this is kind of... like these three cars stand out cause they're all different colors (cars on the foreground). But then this (cars on the background) is like blendy to me. So if I focus on something contrasting (the lady), then my vision seems to be better with like trying to solve the question... It's kind of like a palette cleanser... It's like trying not to second guess the ability of my vision. It brings me more confidence.''} Similarly, when trying to identify the color of a person's shirt in a low contrast area of the image, P4 briefly looked at a higher contrast region with colors similar to the shirt's. This helped her recalibrate her perception and regain confidence in determining the shirt's color.

We also observed visual confusion from low vision participants' gaze behavior during our study. P2 expressed confusion during the study due to misidentification of objects, which led to irregular gaze behaviors to repeatedly examine the object they thought they identified correctly---she thought there was a huge book on the table (Fig \ref{fig:stats}h), but it was actually a laptop with a bookshelf on the background. She had to look up and down to locate the top of the book which caused confusion to her, realizing this was not a book. 


\subsection{Gaze Behavior under Different Visual Intents}
Participants demonstrated diverse gaze behaviors across visual intents shared by both low vision and sighted participants. We report how gaze behaviors differ across visual intents. 

\textit{\textbf{Fixations.}}
We found that \textit{VisualIntent} had significant effects on mean fixation duration (ART: $F_{(4, 133.5)} = 3.83$, $p = 0.006$, $\eta_p^2 = 0.10$). Using a post-hoc contrast test, we found that participants had significantly longer fixations during observing than searching ($t_{(134)} = 3.31$, $p = 0.010$), traversing ($t_{(133)} = 3.15$, $p = 0.017$), and exploring ($t_{(132)} = 3.02$, $p = 0.025$), indicating that during observing, participants had deeper cognitive processing than during the other three visual intents. 
For fixation rate, we observed a trend towards a main effect of \textit{VisualIntent} ($F_{(4, 133.4)} = 2.25$, $p = 0.067$, $\eta_p^2 = 0.063$). However, post-hoc comparisons did not reveal any significant pairwise differences.

\textbf{\textit{Saccades.}} Mean saccade amplitude was significantly affected by \textit{VisualIntent} ($F_{(4, 135.5)} = 28.4$, $p < 0.001$, $\eta_p^2 = 0.46$). A post-hoc comparison suggested that participants had significantly shorter saccades during observing compared to searching ($t_{(136)} = -7.63$, $p < 0.001$), traversing ($t_{(134)} = -5.57$, $p < 0.001$), comparing ($t_{(134)} = -9.46$, $p < 0.001$) and exploring ($t_{(133)} = -8.41$, $p < 0.001$). This finding suggested that the fixations during observing were denser, in line with our qualitative result. We also found that comparing presented longer saccades than traversing ($t_{(134)} = 9.46$, $p < 0.001$). Since saccade amplitudes tend to decrease with increasing cognitive load \cite{ceder1977drivers, may1990eye}, this result suggested potential higher cognitive demands for traversing than comparing.

\textbf{\textit{Stationary Entropy.}} 
We observed a significant main effect of \textit{VisualIntent} on stationary entropy (LME: $\chi^2(4) = 40.1$, $p < 0.001$, $\eta_p^2 = 0.41$, 95\%CI [0.30, 1.0]). Using Tukey's HSD, we found observing had significantly lower entropy than searching ($t_{(136)} = -5.70, p < 0.001$), traversing ($t_{(134)} = -7.36, p < 0.001$), comparing ($t_{(134)} = -6.73, p < 0.001$), and exploring ($t_{(133)} = -8.86, p < 0.001$). This suggested that fixations during observing were more concentrated, whereas fixations during the other four visual intents were more scattered across the whole image. 



\textbf{\textit{Number of Objects Visited.}}
Number of Objects Visited was significantly affected by \textit{VisualIntent} ($F_{(4, 135.3)} = 21.15, p < 0.001, \eta^2_p = 0.38$). Through a post-hoc contrast test for ART, we found that participants visited significantly fewer objects during observing than during searching ($t_{(136)} = -3.50, p = 0.006$), traversing ($t_{(134)} = -8.96, p < 0.001$), comparing ($t_{(134)} = -5.23, p < 0.001$), and exploring ($t_{(133)} = -5.59, p < 0.001$), while traversing covered more objects than searching ($t_{(138)} = 5.15, p < 0.001$), comparing ($t_{(137)} = 3.67, p = 0.003$), and exploring ($t_{(134)} = 3.58, p = 0.004$). 


\textbf{\textit{Attention Distribution over Objects.}}
We discovered a significant main effect of \textit{VisualIntent} on object attention variability ($F_{(4, 134.9)} = 24.7, p < 0.001, \eta^2_p = 0.42$).
Through post-hoc comparisons, we found that observing led to significantly higher object attention variability than searching ($t_{(136)} = 7.31, p < 0.001$), traversing ($t_{(134)} = 8.24, p < 0.001$), comparing ($t_{(134)} = 5.25, p < 0.001$), and exploring ($t_{(133)} = 8.39, p < 0.001$), signifying that some objects received significantly more attention than others during observing. Additionally, comparing had significantly higher object attention than traversing ($t_{(136)} = 2.95, p = 0.030$) and exploring ($t_{(134)} = 2.94, p = 0.031$). This finding corresponds to the fact that comparing involves frequent attention switching between certain objects, whereas traversing and exploring allocate attention to each object more evenly.

We also found \textit{VisualIntent} significantly affected the proportion of time participants spent on foreground objects ($F_{(4, 133.8)} = 6.34, p < 0.001, \eta^2_p = 0.16$).
A post-hoc contrast test revealed that participants spent significantly less time on foreground objects during searching than during observing ($t_{(135)} = -3.23, p = 0.013$), traversing ($t_{(136)} = -3.13, p = 0.018$), and comparing ($t_{(135)} = -3.12, p = 0.018$). Similarly, 
participants showed smaller foreground attention ratio during exploring than during observing ($t_{(132)} = -3.40, p = 0.008$), traversing ($t_{(133)} = -3.30, p = 0.011$) and comparing ($t_{(133)} = -3.28, p = 0.011$). This result revealed that participants attended to background more during searching and exploring than other visual intents.

\subsection{Comparing Low Vision and Sighted People's Gaze Behavior}
We compared how low vision and sighted participants' gaze behavior differed under different visual intents.

\textbf{\textit{Stationary Entropy.}} 
While no significant interaction between \textit{VisualIntent} and \textit{Vision} was found for fixation and saccade measures, we did observe a significant interaction on stationary entropy ($\chi^2(4) = 16.6$, $p = 0.002$, $\eta_p^2 = 0.11$, 95\% CI [0.02, 1.0]).
Consistent with the main effect of \textit{VisualIntent},  sighted participants showed significantly lower entropy during observing compared to searching ($t_{(135)} = -5.39, p < 0.001$), traversing ($t_{(132)} = -4.51, p < 0.001$), comparing ($t_{(132)} = -4.91, p < 0.001$), and exploring ($t_{(132)} = -4.77, p < 0.001$). Similarly, among low vision participants, entropy during observing was significantly lower than during traversing ($t_{(135)} = -5.87, p < 0.001$), comparing ($t_{(135)} = -4.61, p < 0.001$), and exploring ($t_{(133)} = -7.76, p < 0.001$). Unlike the main effect, we found searching had significantly lower entropy than exploring ($t_{(136)} = -4.54, p < 0.001$).
These results suggest that observing consistently involved more focused visual attention for both groups, while exploring showed more dispersed scan paths than searching for low vision participants.

We also found that sighted participants exhibited lower entropy than low vision participants during traversing ($t_{(159)} = -3.25, p = 0.044$) and exploring ($t_{(155)} = -4.62, p < 0.001$). This might be due to the need for some low vision participants to scan more widely to compensate their peripheral vision loss. Fig \ref{fig:stats}c and \ref{fig:stats}d show an example of the difference between sighted and low vision participants during exploring. Sighted participants usually could receive contextual information through peripheral vision, but participants with peripheral vision loss needed to explore broader area.

\textbf{\textit{Number of Objects Visited.}}
We found a significant main effect of \textit{Vision} ($F_{(1, 37.4)} = 16.2, p < 0.001, \eta^2_p = 0.30$), as well as a significant interaction between \textit{VisualIntent} and \textit{Vision} ($F_{(4, 135.4)} = 6.92, p < 0.001, \eta^2_p = 0.17$) on number of objects visited (NOV).
Overall, low vision participants visited significantly more objects than sighted participants. This might be due to their visual conditions, which required them to scan more objects to gather sufficient contextual information for different visual tasks. For example, fig \ref{fig:stats} a and \ref{fig:stats}b show the difference in NOV P1 and C19 touched on when tasked to describe the event shown on the image---P1 fixated at more objects than C19 in general. 

Post-hoc comparisons further revealed patterns consistent with the main effect of \textit{VisualIntent} in both sighted group and low vision group. Among sighted participants, observing involved fewer objects than searching ($t_{(136)} = -4.09, p = 0.003$), traversing ($t_{(133)} = -5.10, p < 0.001$), comparing ($t_{(133)} = -3.72, p = 0.011$), and exploring ($t_{(133)} = -3.74, p = 0.010$). Similarly, among low vision participants, observing involved fewer objects than traversing ($t_{(137)} = -5.99, p < 0.001$) and exploring ($t_{(133)} = -3.61, p = 0.015$). Notably, we found searching covered significantly fewer objects than traversing ($t_{(142)} = -4.87, p = 0.0001$) in low vision group, possibly due to participants terminating their search as soon as the target was found, whereas traversing required more extensive scanning across objects.

\textbf{\textit{Attention Distribution over Objects.}}
There is a significant interaction between \textit{VisualIntent} and \textit{Vision} group ($F_{(4, 134.7)} = 4.23, p = 0.003, \eta^2_p = 0.11$) on object attention variability, but not on foreground attention ratio. Within the sighted group, observing showed significantly higher variability than searching ($t_{(135)} = 6.95, p < 0.001$), traversing ($t_{(133)} = 5.51, p < 0.001$), and exploring ($t_{(133)} = 5.86, p < 0.001$). Additionally, comparing had significantly higher variability than searching ($t_{(137)} = 4.19, p = 0.002$), indicating that for sighted people, searching resulted in more even attention distribution than comparing.
Among low vision participants, observing resulted in significantly higher variability than traversing ($t_{(135)} = 6.20, p < 0.001$), comparing ($t_{(135)} = 4.72, p < 0.001$), exploring ($t_{(133)} = 6.12, p < 0.001$), and searching ($t_{(137)} = -3.25, p = 0.045$), consistent with the main effect of \textit{VisualIntent} observed across all participants.

\subsection{Gaze Behavior Affected by Different Visual Abilities}
In this section, we report the effect of different visual abilities (i.e., visual acuity, peripheral vision) on low vision participants' gaze behaviors.

\textbf{\textit{Saccades.}} Although we found no significant interaction between \textit{VisualIntent} and visual abilities (\textit{PeripheralVision}, \textit{VisualAcuity}) on fixation measures, we found a significant main effect of \textit{VisualAcuity} on mean saccade amplitude ($F_{(1, 19.4)} = 4.80, p = 0.041, \eta^2_p = 0.20$), where low vision participants with low visual acuity had significantly shorter saccades than those with high visual acuity, consistent with low vision people's gaze behaviors during reading \cite{wang2023understanding}. 
Furthermore, a significant interaction between \textit{VisualIntent} and \textit{VisualAcuity} was found on mean saccade amplitude ($F_{(4, 53.8)} = 3.97, p = 0.007, \eta^2_p = 0.23$). 
Using a post-hoc contrast test, we discovered consistent patterns with main effect of \textit{VisualIntent} for low visual acuity group---low vision participants with low visual acuity had significantly shorter saccades during observing than searching ($t_{(57.4)} = -3.79, p = 0.012$), comparing ($t_{(54.0)} = -3.64, p = 0.020$), and exploring ($t_{(52.3)} = -4.85, p < 0.001$). For those with high visual acuity, observing also involved significantly shorter saccades than exploring ($t_{(52.5)} = -5.73, p < 0.001$), and traversing involved shorter saccades than exploring ($t_{(52.5)} = -3.31, p = 0.050$). Though no significant difference was observed between low visual acuity group and high visual acuity group on the same visual intent, using Spearman's correlation test, we found a positive correlation between participants visual acuity (better eye) and saccade amplitude during observing ($r(16) = 0.48, p = 0.046$). This suggested low vision participants with low visual acuity tended to have short saccade length, indicating higher cognitive load\cite{ceder1977drivers, may1990eye}. 

Additionally, a significant main effect of \textit{PeripheralVision} was also observed ($F_{(1, 18.9)} = 6.59, p = 0.019, \eta^2_p = 0.26$)---participants who had peripheral vision loss (according to our visual field test) exhibited shorter saccades than those with intact peripheral vision, echoing low vision people's visual experience in reading \cite{wang2023understanding}. Since peripheral vision loss made it difficult to track where they are on the image, participants with peripheral vision loss tended to move carefully to make sure they gathered all information. 

\textbf{\textit{Number of Objects Visited.}} We looked into how different visual abilities and visual intent affected low vision participants' object-level gaze behavior and found an significant main effect of \textit{PeripheralVision} on the number of objects participants attended to ($F_{(1, 19.3)} = 5.46, p = 0.030, \eta^2_p = 0.22$). On average, participants with peripheral vision loss visited more objects than those with intact peripheral vision, suggesting that reduced peripheral vision might lead to broader visual scanning to gather sufficient contextual information. Fig \ref{fig:stats}e and \ref{fig:stats}f reveal the impact of peripheral vision loss. Unlike P11 with better peripheral vision, P7 had to briefly search in the left side of the image (sofa) to make sure no other person is in the image due to his reduced field of view. 

No significant main effect of visual abilities (\textit{PeripheralVision}, \textit{VisualAcuity}) or their interaction with \textit{VisualIntent} was found on stationary entropy, object attention variability, and foreground attention ratio. 

%% file: sections/5-Discussion.tex
\section{Discussion}
To inspire future intent-aware assistive technology for low vision users, comprehensively understanding low vision people's visual intent is the essential first step. To this end, we contributed the first research that investigated visual intent in both low vision and sighted people. Unlike prior literature on visual intent recognition that employed predefined visual intents for specific tasks and context, we adopted a bottom-up approach---eye tracking-based retrospective think-aloud---to thoroughly capture people's visual intents both qualitatively and quantitatively via diverse image-viewing tasks. By analyzing participants' gaze recordings alongside their subjective visual experiences, we built a visual intent taxonomy that includes five categories---searching, observing, traversing, comparing and exploring---shared by both low vision and sighted participants. Additionally, we identified unique goals and gaze behaviors among low vision participants, such as recalibrating color perception, beyond the five visual intents.

Comparing participants' gaze behavior in different visual intents, we found observing consistently stood out, characterized by longer fixation duration, shorter saccades, lower stationary entropy, and higher object attention variability---indicating more focused attention on specific objects. Besides, we found gaze behavior under comparing showed longer saccades than traversing, and more uneven attention distribution than both traversing and exploring, reflecting the need to switch between reference targets. Moreover, we found searching and exploring involved more time spent on image backgrounds, suggesting broader, less focused scanning strategies.
Between-group comparisons revealed that low vision participants scanned more broadly than sighted participants during traversing and exploring, likely due to reduced peripheral vision. 
Furthermore, our analysis of visual ability among low vision participants showed that both low visual acuity and peripheral vision loss were associated with shorter saccades---higher cognitive load, while peripheral vision loss also led to a greater number of objects being visited---indicating compensatory scanning to gather contextual information.

Our study provides rich insights into the gaze behaviors and visual experiences of low vision users, including visual challenges during image-viewing tasks. These findings can inform the development of visual intent recognition pipelines tailored to low vision users and guide the design of gaze-based interactive systems that provide timely, intent-aware support in daily tasks. In the following discussion, we reflect on our data collection methods, the generalizability of our visual intent taxonomy, propose directions for intent recognition approaches, and outline design implications for intent-aware assistive technologies for low vision users.

\subsection{Gaze Data Collection for Model Training}
As the first study that employed eye-tracking based retrospective think-aloud (ETRTA) method with low vision users, we were initially concerned about whether their visual conditions would limit their ability to recall gaze behaviors during each trial. However, both sighted and low vision participants were able to clearly verbalize their gaze trajectory and intent for each movement before we showed them their gaze recordings, demonstrating the feasibility of our method. The validity of our method was further supported by our quantitative findings, where visual intent had significant effects on multiple gaze measures. These results suggest that ETRTA is a suitable approach for constructing a visual intent taxonomy.

However, ETRTA poses limitations for future data collection for intent recognition models. Because participants' gaze recordings were segmented based on their description and clarification, there is no control over how many segments they produce or how long each segment is. Since most intent recognition models \cite{hu2021ehtask, wang2024tasks} utilize windowing strategies---cutting gaze recording to segments with the same length---for model input, ETRTA might not be the optimal method.
Additionally, ETRTA is time-intensive, making it less practical for large-scale dataset construction. Future research should explore more efficient and standardized methods for collecting intent-labeled gaze data from low vision users. For example, Wang et al. \cite{wang2024tasks} developed a VR-based protocol in which participants completed single-intent visual tasks with pauses in between, allowing for standardized segment labeling while giving users time to reposition and prepare for the next task.

\subsection{Generalizability of Visual Intent Taxonomy}
The visual intent taxonomy we derived was based on diverse image-viewing tasks on a 2D screen. 
While we ensured accessibility by enlarging images to fill the entire screen, our study did not involve assistive tools such as screen magnifiers or contrast enhancement---features commonly used by people with low vision on modern computers.
When using these assistive tools, the user's visual intents may alter due to the changes in how they interact with visual content, which can subsequently affect their gaze behavior. For example, when navigating a computer screen with a screen magnifier, users may engage in more smooth pursuits to follow the magnified window, leading to different gaze patterns compared to viewing static, full-screen images without magnification.

Extending our visual intent taxonomy to more dynamic visual scenarios (e.g., video watching, real-world activities) should be further explored, as the unique affordances in these contexts---the emergence and movements of visual elements---can elicit new visual intents and motion-induced gaze behaviors, such as tracking moving objects\cite{wang2024tasks}, and predicting their trajectories \cite{tuhkanen2021visual}. Moreover, visual tasks in real-world 3D scenes also involve head movements and varying gaze depth (vergence) \cite{mahanama2022eye}, which are not observed in 2D screen-based settings.

Future research should investigate how visual intents manifest under different assistive settings and dynamic viewing contexts to assess whether new intents emerge, how existing intents evolve, and how gaze-based intent recognition systems can be adapted accordingly.

\subsection{Context-Aware Visual Intent Recognition Based on Visual Abilities}
Our findings suggested that both visual acuity and peripheral vision can affect a low vision user's gaze behaviors in different visual intents. 
Therefore, existing machine learning based visual intent recognition systems designed for sighted users \cite{hu2021ehtask, wang2024tasks} might not work well for low vision users. 
To make visual intent recognition more accessible, visual ability information should be incorporated into AI models. Moreover, we found visual context is crucial in interpreting visual intent which can be combined with visual ability to better assist the recognition of visual intent for scenario-specific tasks.

While visual intent had significant effect on participants' gaze behaviors, differences among searching, traversing and exploring were generally not significant across most measures. This result indicates that these visual intents were inherently difficult to distinguish, which again highlights the necessity of incorporating context information to assist intent recognition. Moreover, gaze and intent history might also facilitate the intent recognition of similar gaze behaviors. Recent advancement of multimodal large language model (MLLM) has shown great potential in supporting daily tasks by integrating gaze and context information\cite{boiarshinov2024providing, rekimoto2025gazellm, zhang2025mindeye}. For example, Rekimoto \cite{rekimoto2025gazellm} proposed an MLLM-based assistant using 1st-person video frames of cooking and gaze positions as input to provide accurate cooking guidance. Future research should consider applying MLLMs to the analysis of gaze data to facilitate the interpretation and recognition of visual intent for low vision users.

\subsection{Intent-Aware Support for Low Vision}
Our research highlights the potential of intent-aware augmentation to address low vision people's visual challenges. Based on visual intent recognition result and visual context, we can design various augmentations to support specific visual intent. For example, in a screen-based application, when comparing is detected, the system can selectively enhance the relevant objects using magnification or high-contrast contours to support more efficient comparisons. Future work can also explore deploying such intent-aware support in real-world environments using wearable eye trackers embedded in AR (augmented reality) devices. For instance, when observing is detected as a user fixates on a traffic light, the system can visually magnify it or provide audio feedback to support timely and informed decision-making. Building on the visual needs and challenges reported by our low vision participants, we propose the following directions for the design of future intent-aware low vision support:

\subsubsection{Visual Confusion Support}
Although not formally categorized as a visual intent, visual confusion occurs when low vision users misidentify an object and realize that it does not match their expectation (e.g., P2). Such confusion state can be recognized when gaze behavior does not align with the typical patterns seen in known visual intents.
When confusion is detected, future assistive technology should not only apply general object-level augmentations but also prompt the user to clarify the cause of confusion. These interactions can be used to refine intent recognition models and personalize support based on the user's gaze behaviors and viewing habits. Over time, the system can learn to proactively respond when similar confusion patterns happen again, providing more timely and context-aware assistance.

\subsubsection{Layered Visual Augmentation}
Two participants mentioned the need to recalibrate color perception in order to regain clarity in low-contrast regions of an image. This behavior often occurs across various intents (e.g., comparing and traversing), making it difficult to detect through intent recognition alone. To address this, future technologies should go beyond general object-based enhancement. For example, systems could offer layered visual augmentation at multiple levels: low-level support could target specific visual features such as color characteristics in low-contrast regions; mid-level support might account for the number and types of affected objects; and high-level support could incorporate spatial and semantic relationships or interactions among those objects. Such nuanced support would allow users to engage with visual augmentation at their preferred level, helping them complete visual tasks more confidently and efficiently without second-guessing their perception.

\subsubsection{Peripheral Vision Support}
According to our findings, peripheral vision loss significantly changed a person's gaze behavior, causing difficulty in various visual tasks. However, we found that their gaze behavior was correlated with the boundary of their visual field. To better support users with peripheral vision loss, future technology can consider forecasting what is just outside the user's current visual field when their eye gazes show a tendency to move towards the boundary of their current visual field by rendering the outside objects within the user's current field of view. While it is true that the visual field shifts with eye movement, identifying moments as their gaze reaches the visual field boundary could enable the system to anticipate users' next steps. This predictive augmentation can potentially reduce the need to scan excessively.

\subsection{Limitations}
Our research has several limitations. Despite the challenges of recruiting low vision participants with specific visual conditions, our study included 9 participants with low visual acuity and 11 with peripheral vision loss. However, 6 participants had both conditions, resulting in overlap between the groups. This reduced the independence of our visual ability categories and limited the statistical power of our analysis. Therefore, although our data showed no significant interaction effect between low visual acuity and peripheral vision loss on participants' gaze behavior (beyond the combined effects of each), this result should be interpreted with caution.  Future work should consider involving larger and more balanced samples to better isolate the effects of different types of visual abilities and further improve the generalizability of our findings.

\section{Conclusion}

Our work contributed the first investigation of visual intent categorization with both low vision and sighted people, using a bottom-up, eye-tracking-based retrospective think-aloud method. With 20 low vision and 20 sighted participants, we developed a visual intent taxonomy including searching, observing, traversing, comparing, and exploring, shared by both low vision and sighted participants, and uncovered low vision–specific gaze behaviors. Our analysis revealed distinct gaze behaviors under each visual intent, as well as differences between sighted and low vision users, and differences shaped by visual abilities like low visual acuity and peripheral vision loss. These findings provide a comprehensive understanding of visual intent for low vision users and inform future directions in visual intent recognition and intent-aware low vision assistive technology.

%% file: sections/appendix.tex
\section{Appendix}
\subsection{Questions for Image-Viewing Tasks}
Below are the questions used in our study; repeated questions across images are listed only once.
\label{questionlist}
\subsubsection{Level 1: Within-Object Information}
\begin{questionlist}
    \item How to get to the T1 Departures?
    \item How is the weather?
    \item Is there a person wearing a green plaid shirt?
    \item What accessories does the woman wear?
    \item What are the students holding?
    \item What color are the woman's gloves?
    \item What color is the backpack worn by the boy on the right?
    \item What color is the big truck?
    \item What color is the boy's hair?
    \item What color is the cap of the tallest boy?
    \item What color is the cap the man in the middle wears?
    \item What color is the car on the left of the image?
    \item What color is the car on the right?
    \item What color is the chair on the left?
    \item What color is the coat the mannequin wears on the right of the image?
    \item What color is the crossbody bag on the left?
    \item What color is the dog?
    \item What color is the dress the girl is wearing in the center?
    \item What color is the egg tart?
    \item What color is the girl's backpack?
    \item What color is the girl's clothes?
    \item What color is the helmet of the biggest person?
    \item What color is the ice cream?
    \item What color is the kid's t-shirt?
    \item What color is the man's beard?
    \item What color is the man's headscarf?
    \item What color is the man's shirt?
    \item What color is the mask worn by the woman on the right in this image?
    \item What color is the sofa in the center of the image?
    \item What color is the speaker's t-shirt?
    \item What color is the suit of the person in front of the microphone?
    \item What color is the swimming cap on the top of the image?
    \item What color is the t-shirt of the jumping man?
    \item What color is the t-shirt of the person standing on the right?
    \item What color is the t-shirt the man wears on the left of this photo?
    \item What color is the tablet?
    \item What color is the tent?
    \item What color is the towel hanging on the wall?
    \item What color is the umbrella?
    \item What color is the washer on the left?
    \item What color is the woman in the center wearing?
    \item What color is the woman's backpack?
    \item What color is the woman's clothes?
    \item What color is the woman's dress?
    \item What color is the woman's pants?
    \item What color is the woman's t-shirt?
    \item What color of clothes does the person on the left wear?
    \item What color t-shirt does the biggest girl wear?
    \item What food is in the center of the table?
    \item What food is inside the bowl?
    \item What food is the person cooking?
    \item What ingredient is on the biggest donut?
    \item What is his facial expression?
    \item What is in the bowl on the right?
    \item What is in the middle of the image?
    \item What is in the pan on the left?
    \item What is next to the woman's laptop?
    \item What is on the blue door on the left?
    \item What is on the face of the person on the left?
    \item What is on the plate?
    \item What is on the table?
    \item What is on the wall?
    \item What is the biggest person holding?
    \item What is the color of the banner?
    \item What is the color of the woman's nails?
    \item What is the facial expression of the man leaning on the couch?
    \item What is the facial expression of the man on the left?
    \item What is the facial expression of the person on the right?
    \item What is the facial expression of the person wearing pink?
    \item What is the facial expression of the woman on the left?
    \item What is the facial expression of the woman?
    \item What is the food on the plate next to the mug?
    \item What is the hair color of the woman in the center?
    \item What is the hairstyle of the girl on the left?
    \item What is the hand gesture of the person in the gray t-shirt on the left?
    \item What is the man holding?
    \item What is the man wearing on his head?
    \item What is the man's facial expression?
    \item What is the number at the bottom right of the image?
    \item What is the pattern of the carpet?
    \item What is the person doing?
    \item What is the person in the red shirt on the right doing?
    \item What is the person on the left holding?
    \item What is the posture of the girl on the right?
    \item What is the posture of the man on the left?
    \item What is the posture of the woman?
    \item What is the title of the document on this screenshot?
    \item What is the woman holding?
    \item What is the woman in yellow holding?
    \item What is the woman's facial expression?
    \item What is the woman's hair color?
    \item What kind of car is in the image?
    \item What kind of food is on the plate?
    \item What kind of instrument is the man on the front playing?
    \item What kind of instrument is the person on the left playing?
    \item Where is the baby?
    \item Where is the calendar application?
    \item Where is the search bar?
    \item Where is the snack with pink packaging?
    \item Which folder is selected on this google drive page in the screenshot?
    \item Who is wearing eye glasses?
    \item Who is wearing sneakers?
    \item Who is wearing the blue sneakers?
\end{questionlist}


\subsubsection{Level 2: Cross-Object Information}

\begin{questionlist}
    \item How many boats are there in this photo?
    \item How many buses are shown in this photo?
    \item How many cars are there in this photo?
    \item How many chairs are there?
    \item How many cups are there?
    \item How many donuts are there?
    \item How many drawers are there?
    \item How many egg tarts are there?
    \item How many flavors of ice cream are there?
    \item How many icons are there in this screenshot?
    \item How many mannequins are there?
    \item How many open drawers are in this photo?
    \item How many pans are there on the stovetop?
    \item How many people are crossing the street?
    \item How many people are facing the sea?
    \item How many people are in the hallway?
    \item How many people are in this photo?
    \item How many people are not standing?
    \item How many people are raising their arms?
    \item How many people are receiving the awards?
    \item How many people are seated?
    \item How many people are standing next to the speaker?
    \item How many people are there in the second row?
    \item How many people are wearing sunglasses?
    \item How many people wear the green apron?
    \item How many pillows are on the bed?
    \item How many pineapples are on the shelf?
    \item How many plates are there on the table?
    \item How many police cars are there?
    \item How many policemen are there?
    \item How many pots are there on the stovetop?
    \item How many sinks are there in this photo?
    \item How many snacks are there?
    \item How many suitcases are in this photo?
    \item How many tabs are open in the chrome browser?
    \item How many throw pillows are there?
    \item How many toy cars are there?
    \item What are people on the right of this photo doing?
    \item What are the two boys in the middle of the photo doing?
    \item What are the two girls doing?
    \item What are the two people doing?
    \item What are the two people in the middle doing?
    \item What are the two women on the left doing?
    \item What are these three people doing?
    \item What are people on the image doing?
    \item What food is the person cooking?
    \item What is the man doing?
    \item What is the man looking at?
    \item What is the person doing with their hands?
    \item What is the woman and the little kid doing?
    \item What is the woman doing?
    \item What is this person holding?
    \item What is on the table?
    \item What is the relationship between the people in this image?
    \item Who is the man on the left looking at?
\end{questionlist}

\vspace{1em}

\subsubsection{Level 3: Overall Interpretation Information}

\begin{questionlist}
    \item What activity does this image describe?
    \item What are the characteristics of this alley?
    \item What emotion does the image convey?
    \item What event does the image describe?
    \item What event might have happened before this picture was taken?
    \item What is the atmosphere of this image?
    \item What is the purpose of taking this photo?
    \item What is the purpose of the application shown in the image?
    \item What is the purpose of showing this image?
    \item What kind of store is this?
    \item Where are the egg tarts?
    \item Where is this photo taken?
\end{questionlist}